\documentclass[twocolumn,prl,a4paper,superscriptaddress,showpacs,
preprintnumbers,tightenlines,footnote]{revtex4}
\draft 
\usepackage{amsmath}
\usepackage{graphicx}
\usepackage{epsfig}

\usepackage{dcolumn}
\newcolumntype{B}{D{B}{}{-1}}

\begin{document}

\setlength{\unitlength}{1pt}

\title{ \quad\\[1.0cm] Measurement of inclusive {\boldmath $D_s$}, 
{\boldmath $D^0$}, and {\boldmath $J/\psi$} rates
and determination of the {\boldmath $B_s^{(\ast)} \bar{B}_s^{(\ast)}$}
production fraction in {\boldmath $b \bar{b}$}
events at the {\boldmath $\Upsilon$(5S)} resonance}

\affiliation{Budker Institute of Nuclear Physics, Novosibirsk}
\affiliation{Chiba University, Chiba}
\affiliation{Chonnam National University, Kwangju}
\affiliation{University of Cincinnati, Cincinnati, Ohio 45221}
\affiliation{University of Frankfurt, Frankfurt}
\affiliation{The Graduate University for Advanced Studies, Hayama} 
\affiliation{University of Hawaii, Honolulu, Hawaii 96822}
\affiliation{High Energy Accelerator Research Organization (KEK), Tsukuba}
\affiliation{Hiroshima Institute of Technology, Hiroshima}
\affiliation{University of Illinois at Urbana-Champaign, Urbana, Illinois 61801}
\affiliation{Institute of High Energy Physics, Chinese Academy of Sciences, Beijing}
\affiliation{Institute of High Energy Physics, Vienna}
\affiliation{Institute of High Energy Physics, Protvino}
\affiliation{Institute for Theoretical and Experimental Physics, Moscow}
\affiliation{J. Stefan Institute, Ljubljana}
\affiliation{Kanagawa University, Yokohama}
\affiliation{Korea University, Seoul}
\affiliation{Kyungpook National University, Taegu}
\affiliation{Swiss Federal Institute of Technology of Lausanne, EPFL, Lausanne}
\affiliation{University of Ljubljana, Ljubljana}
\affiliation{University of Maribor, Maribor}
\affiliation{University of Melbourne, Victoria}
\affiliation{Nagoya University, Nagoya}
\affiliation{Nara Women's University, Nara}
\affiliation{National Central University, Chung-li}
\affiliation{National United University, Miao Li}
\affiliation{Department of Physics, National Taiwan University, Taipei}
\affiliation{H. Niewodniczanski Institute of Nuclear Physics, Krakow}
\affiliation{Nippon Dental University, Niigata}
\affiliation{Niigata University, Niigata}
\affiliation{University of Nova Gorica, Nova Gorica}
\affiliation{Osaka City University, Osaka}
\affiliation{Osaka University, Osaka}
\affiliation{Panjab University, Chandigarh}
\affiliation{Peking University, Beijing}
\affiliation{Princeton University, Princeton, New Jersey 08544}
\affiliation{RIKEN BNL Research Center, Upton, New York 11973}
\affiliation{Saga University, Saga}
\affiliation{University of Science and Technology of China, Hefei}
\affiliation{Seoul National University, Seoul}
\affiliation{Shinshu University, Nagano}
\affiliation{Sungkyunkwan University, Suwon}
\affiliation{University of Sydney, Sydney, New South Wales}
\affiliation{Tata Institute of Fundamental Research, Bombay}
\affiliation{Toho University, Funabashi}
\affiliation{Tohoku Gakuin University, Tagajo}
\affiliation{Tohoku University, Sendai}
\affiliation{Department of Physics, University of Tokyo, Tokyo}
\affiliation{Tokyo Institute of Technology, Tokyo}
\affiliation{Tokyo Metropolitan University, Tokyo}
\affiliation{Tokyo University of Agriculture and Technology, Tokyo}
\affiliation{Virginia Polytechnic Institute and State University, Blacksburg, Virginia 24061}
\affiliation{Yonsei University, Seoul}
  \author{A.~Drutskoy}\affiliation{University of Cincinnati, Cincinnati, Ohio 45221} 
  \author{K.~Abe}\affiliation{High Energy Accelerator Research Organization (KEK), Tsukuba} 
  \author{K.~Abe}\affiliation{Tohoku Gakuin University, Tagajo} 
  \author{I.~Adachi}\affiliation{High Energy Accelerator Research Organization (KEK), Tsukuba} 
  \author{H.~Aihara}\affiliation{Department of Physics, University of Tokyo, Tokyo} 
  \author{D.~Anipko}\affiliation{Budker Institute of Nuclear Physics, Novosibirsk} 
  \author{K.~Arinstein}\affiliation{Budker Institute of Nuclear Physics, Novosibirsk} 
  \author{V.~Aulchenko}\affiliation{Budker Institute of Nuclear Physics, Novosibirsk} 
  \author{T.~Aushev}\affiliation{Swiss Federal Institute of Technology of Lausanne, EPFL, Lausanne}\affiliation{Institute for Theoretical and Experimental Physics, Moscow} 
  \author{S.~Banerjee}\affiliation{Tata Institute of Fundamental Research, Bombay} 
% ...
\collaboration{The Belle Collaboration}

\date{\today}

\tighten

\begin{abstract}
The inclusive production of $D_s$, $D^0$, and $J/\psi$ mesons is studied
using a 1.86\,fb$^{-1}$ data sample collected
on the $\Upsilon$(5S) resonance with the Belle 
detector at the KEKB asymmetric energy $e^+ e^-$ collider.
The number of $b\bar{b}$ events in this $\Upsilon$(5S)
data sample is determined.
We measure the branching fractions
\mbox{${\cal B}(\Upsilon({\rm 5S})\rightarrow D_s X)\,/\,2 =$}
\mbox{$(23.6 \pm 1.2 \pm 3.6)\%$},
\mbox{${\cal B}(\Upsilon({\rm 5S})\rightarrow D^0 X) / 2 =$}
\mbox{$(53.8 \pm 2.0 \pm 3.4)\%$}, and
\mbox{${\cal B}(\Upsilon({\rm 5S})\rightarrow J/\psi X) / 2 =$}
\mbox{$(1.030 \pm 0.080 \pm 0.067)\%$}.
From the $D_s$ and $D^0$ inclusive branching fractions the ratio 
\mbox{$f_s = (18.0 \pm 1.3 \pm 3.2)\%$} of 
$B_s^{(\ast)} \bar{B}_s^{(\ast)}$ to the total
$b\bar{b}$ quark pair production at the $\Upsilon$(5S) energy
is obtained in a model-dependent way.
\end{abstract}


\maketitle

\tighten

{\renewcommand{\thefootnote}{\fnsymbol{footnote}}}
\setcounter{footnote}{0}

The possibility of studying $B_s$ decays at very high luminosity
$e^+ e^-$ colliders running at the energy of the $\Upsilon$(5S) resonance
has been discussed in several theoretical papers \cite{teoa,teob,teoc}.
Studies of the $B_s$ meson properties at the $\Upsilon$(5S),
both alone and in comparison with 
those of its lighter cousins $B^0$ and $B^+$, may provide 
important insights into the Cabibbo-Kobayashi-Maskawa matrix and 
hadronic structure, as 
well as sensitivity to new physics phenomena \cite{npsa}.
To date, most studies of $B_s$ have been performed at
hadron colliders, where high production rates are tempered
by limited triggering and detection capabilities.
As $B$ factories have amply demonstrated for the $B^0$ and $B^+$, the 
kinematic cleanliness of resonant near-threshold exclusive pair 
production ($e^+e^- \to \Upsilon{\rm (4S)} \to B\bar B$), combined with a 
high triggering efficiency and the ability to measure neutral 
particles, can open up a complementary realm of sensitivity to new 
phenomena.
The $\Upsilon$(5S) could play a similar role 
for $B_s$ that the $\Upsilon$(4S) has played for $B$.
 
To test the experimental feasibility of $B_s$ studies in 
$\Upsilon$(5S) events, a sample of 1.86~fb$^{-1}$ was 
collected with the Belle detector
over 3 days in June 2005.
Earlier an $\Upsilon$(5S) data sample of $\sim$0.1~fb$^{-1}$
was taken at CESR \cite{cleol,cusba,cusbb} and, more recently,
a dataset of 0.42~fb$^{-1}$ was collected
by CLEO \cite{cleoi,cleoe}.  
We report here the first results obtained by Belle, 
a determination of the number of $b\bar{b}$ events in the $\Upsilon$(5S)
data sample,
a measurement of 
the inclusive rate of $\Upsilon$(5S) events to $D_s$, $D^0$, and 
$J/\psi$ and derivation of the fraction of $b\bar{b}$ events 
containing $B_s$.

The Belle detector \cite{belle} has operated since 1999 at KEKB \cite{kekb},
an asymmetric-energy double storage ring designed to collide 
8 GeV electrons and 3.5 GeV positrons and produce 
$\Upsilon$(4S) mesons with a boost of $\beta \gamma$ = 0.425.
The recent data sample of $1.86\,\mathrm{fb}^{-1}$ was taken at the
$\Upsilon$(5S) energy of $\sim$10869 MeV, under exactly the same 
experimental conditions as in $\Upsilon$(4S) and continuum runs, 
except that both beam energies were increased by $\sim$2.7$\%$, 
keeping the center-of-mass (CM) boost unchanged. Another data
sample of $3.67\,\mathrm{fb}^{-1}$ collected
at a CM energy \mbox{60\ MeV} below the $\Upsilon$(4S)
just before the $\Upsilon$(5S) data taking
is used in this analysis to evaluate continuum contributions.

Only clean decay modes $D_s^+\,\to\,\phi \pi^+$ ($\phi \to K^+ K^-$), 
$D^0\,\to\,K^- \pi^+$ and $J/\psi\,\to\,\mu^+ \mu^-$ are reconstructed.
Charge-conjugate modes are implicitly included everywhere in this Letter.
The standard track reconstruction and particle identification
procedures are used \cite{belle}.
The invariant mass of $\phi\,\to\,K^+ K^-$ candidates is
required to be within \mbox{$\pm$ 12\,MeV/$c^2$} of the nominal $\phi$ mass.
For the $D_s^+\,\to\,\phi \pi^+$ decay mode, the helicity angle 
distribution is expected to be proportional to 
\mbox{cos$^2 \theta_{\rm hel}^{D_s}$}; therefore,
the requirement \mbox{$|$cos $\theta_{\rm hel}^{D_s}| > 0.25$} is applied.
The helicity angle $\theta_{\rm hel}^{D_s}$ is defined 
as the angle between the directions of the $K^+$ and $D_s^+$ momenta 
in the $\phi$ rest frame.

In the energy region of the $\Upsilon$(5S), hadronic events
can be classified into three physics categories:
$u\bar{u}, d\bar{d}, s\bar{s}, c\bar{c}$ continuum events, 
$b\bar{b}$ continuum events, and $\Upsilon$(5S)
events. The $b\bar{b}$ continuum and the $\Upsilon$(5S) 
events always produce final states with a pair of $B$ or 
$B_s$ mesons and, therefore, cannot be topologically separated.
We define the $b\bar{b}$ continuum and
$\Upsilon$(5S) events collectively
as $b\bar{b}$ events, everywhere in this analysis.
All $b\bar{b}$ events are expected to hadronize
in one of the following final states: $B\bar{B}$,
$B\bar{B}^\ast$, $B^\ast\bar{B}$, $B^\ast\bar{B}^\ast$, 
$B\bar{B}\,\pi$, $B\bar{B}^\ast\,\pi$, $B^\ast\bar{B}\,\pi$, 
$B^\ast\bar{B}^\ast\,\pi$, $B\bar{B}\,\pi \pi$,
$B_s^0\bar{B}_s^0$, $B_s^0\bar{B}_s^\ast$, $B_s^\ast\bar{B}_s^0$ or 
$B_s^\ast\bar{B}_s^\ast$.
Here $B$ denotes a $B^0$ or a $B^+$ meson and 
$\bar{B}$ denotes a $\bar{B}^0$ or a $B^-$ meson.
The excited states decay to their ground states via $B^* \to B \gamma$ and
$B_s^* \to B_s^0 \gamma$ \cite{pdg}.

An energy scan was performed just before the $\Upsilon$(5S) data taking
to find the peak position of the $\Upsilon$(5S) resonance.
An integrated luminosity of $\sim$\,30\,pb$^{-1}$ was collected
at five values of $e^+ e^-$ CM energy between 10825~MeV and 10905~MeV
at intervals of 20~MeV.
The ratio of the number of hadronic events with 
$R_2 > 0.2$ to the number of Bhabha events is measured
as a function of the CM energy (Fig.\ 1).
Here, $R_2$ is the normalized second Fox-Wolfram moment \cite{fox}. 
This ratio is expected
to have a Breit-Wigner function shape corresponding 
to the $\Upsilon$(5S) resonance, above a flat background.
Fixing the width value to the PDG value
$\Gamma =$ \mbox{110 MeV/$c^2$} \cite{pdg},
the mean mass value is found from the fit 
to be $M = (10868 \pm 6 \pm 14)\,$MeV/$c^2$, where the first error
is statistical and the second error is a systematic uncertainty, 
dominated by the variation of background contributions with CM energy.
This value is in good agreement with the PDG value
$M_{\Upsilon{\rm (5S)}} = (10865 \pm 8)\,$MeV/$c^2$ \cite{pdg}.
The fit value obtained above is treated only as a systematic check
rather than as a measurement, as the energy range scanned is small compared 
to the $\Upsilon$(5S) width,
and uncertainties due to background contributions are not well known.
Finally, the energy of 10869~MeV was set for subsequent $\Upsilon$(5S) runs.

\begin{figure}[t!]
\epsfig{file=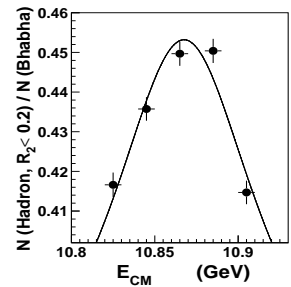,width=3.9cm,height=3.9cm}
\vspace{-0.2cm}
\caption{The ratio of the number of hadronic events with
$R_2 > 0.2$ to the number of Bhabha events,
as a function of the $e^+ e^-$ CM energy.
Only statistical errors are shown. 
The curve is the result of the fit described in the text.}
\vspace{-0.2cm}
\label{fig1}
\end{figure}

The $u\bar{u}, d\bar{d}, s\bar{s}, c\bar{c}$ continuum subtraction 
method is applied to obtain
the number of $b\bar{b}$ events in the $\Upsilon$(5S) data sample:
\begin{equation}
N^{b\bar{b}}_{\rm 5S}=\frac{1}{\epsilon^{b\bar{b}}_{\rm 5S}} \left( N^{\rm had}_{\rm 5S}-N^{\rm had}_{\rm cont}\cdot\frac{{\cal L}_{\rm 5S}}{{\cal L}_{\rm cont}}\cdot\frac{E_{\rm cont}^{\,2}}{E_{\rm 5S}^{\,2}}\cdot\frac{\epsilon^{\rm con}_{\rm 5S}}{\epsilon^{\rm con}_{\rm cont}} \right).
\end{equation}
Here $N^{b\bar{b}}_{\rm 5S}$ is the number of $b\bar{b}$
events in the $\Upsilon$(5S) data sample, and $N^{\rm had}_{\rm 5S}$
and $N^{\rm had}_{\rm cont}$ are the numbers of hadronic events 
in the $\Upsilon$(5S) and continuum data samples, respectively.
A few percent contribution of $\tau^+ \tau^-$, QED, $\gamma \gamma$ 
and beam-gas events partially cancels in Eq.\ (1), and 
the corresponding small systematic uncertainty is included in the 
full systematic uncertainty.
The efficiency to select a $b\bar{b}$ event in the $\Upsilon$(5S) data 
sample, $\epsilon^{b\bar{b}}_{\rm 5S} = (99 \pm 1)\%$, and
the efficiency ratio for continuum events in the $\Upsilon$(5S) 
and continuum data samples, 
$\epsilon^{\rm con}_{\rm 5S} / \epsilon^{\rm con}_{\rm cont} = 1.007 \pm 0.003$,
are obtained from Monte Carlo (MC) simulation.
The hadronic cross section varies with the CM energy 
as $1/E^2$ and a corresponding correction is applied.
The CM energies for the $\Upsilon$(5S) and 
continuum data sets 
are $E_{\rm 5S} = 10869\,$MeV and $E_{\rm cont} = 10520\,$MeV,
respectively, with a $\sim5\,$MeV accuracy of 
the collider absolute CM energy calibration.
The integrated luminosity ratio 
${\cal L}_{\rm 5S} / {\cal L}_{\rm cont} = 0.5061 \pm 0.0020$ 
is calculated using the standard Belle luminosity measurement procedure
with Bhabha events. 
The small statistical uncertainty on this ratio can be neglected.

The value of and uncertainty on the luminosity ratio is further checked
using high-momentum charged tracks, $K^0_S$ mesons, and $D^0$ mesons.
To compare $\Upsilon$(5S)
and continuum production, normalized momentum 
distributions are used. The normalized momentum of a particle $h$
is defined as $x(h) = P(h) / P_{\rm max}(h)$, where
$P(h)$ is the measured momentum of that particle,
and $P_{\rm max}(h)$ is the expected value of its momentum if it were
produced in the process $e^+ e^- \to h \bar{h}$ at the same 
CM energy.
Fitting a constant to the distributions
of the $\Upsilon$(5S) and continuum dataset ratios
$x(h)_{\rm 5S} / x(h)_{\rm cont}$
for $0.5\,<\,x(h)\,<\,0.9$ and applying small corrections
due to the difference between the $\Upsilon$(5S) and continuum 
final state particle multiplicities,
which were obtained from MC simulation,
the ratios $0.471 \pm 0.005$, $0.471 \pm 0.005$, and 
$0.477 \pm 0.005$ are determined for $h = \pi^+, K^0_S$, and $D^0$,
respectively.
These values agree with the energy-corrected luminosity ratio
of $0.4740 \pm 0.0019$ obtained from Bhabha event measurements,
where the factor $E_{\rm cont}^{\,2} / E_{\rm 5S}^{\,2}$ was applied 
to correct for the hadronic cross-section energy dependence.

From Eq.\ (1) we obtain the number of $b\bar{b}$ events
in the $\Upsilon$(5S) data sample,
${\rm N}^{b\bar{b}}_{\rm 5S} = (5.61 \pm 0.03_{\rm stat} \pm 0.29_{\rm syst}) \times 10^5$.
The total systematic uncertainty of \mbox{$\sim 5\%$} includes all 
systematic errors on parameters used
in Eq.\ (1).
Finally, the $b\bar{b}$ production cross-section at the $\Upsilon$(5S) is 
measured to be $(0.302 \pm 0.015)\,$nb,
in good agreement with the CLEO value of 
\mbox{$(0.310 \pm 0.052)\,$nb} \cite{cleoi}.

The method for inclusive $D_s$ ($D_s \equiv D_s^{\pm}$) analysis 
of $\Upsilon$(5S) data developed in Ref.~\cite{cleoi} is applied here.
The $D_s$ signals ($D_s^+ \to \phi \pi^+ , \phi \to K^+ K^-$)
in the $\Upsilon$(5S) and continuum
data samples are shown in Fig.\ 2(a) for the normalized $D_s$ momentum
region $x(D_s) < 0.5$, where a $b\bar{b}$ contribution is expected.
Here and throughout this Letter, the continuum
distributions are normalized to the $\Upsilon$(5S) distributions
using the energy-corrected luminosity ratio.
To extract the number of $D_s$ mesons, the $D_s$ mass distribution is fitted
by a Gaussian to describe the signal and a linear function to describe 
the background. The Gaussian width is fixed to the value obtained from
MC simulation; the Gaussian mean value and normalization, and 
the background parameters are allowed to float.
The same mass fit procedure, but with the Gaussian mean value
fixed to that obtained from the fit
of the $D_s$ signal in the $x(D_s) < 0.5$ range,
is repeated in each bin of $x(D_s)$ in order to obtain the
$D_s$ yield as a function of the normalized
momentum, $x(D_s)$.
The $x(D^0)$ and $x(J/\psi)$ distributions discussed below
are obtained using the same fit procedures.

The normalized momentum $x(D_s)$ distributions are shown in Fig.~2(b)
for the $\Upsilon$(5S) and continuum data samples.
These two distributions agree well in the region $x(D_s) > 0.5$, 
where $b\bar{b}$ events cannot contribute.
The excess of events in the region $x(D_s) < 0.5$ corresponds
to inclusive $D_s$ production in $b\bar{b}$ events.

The fully corrected $x(D_s)$ distribution 
for $b\bar{b}$ events
is obtained, subtracting the continuum contribution and applying 
a bin-by-bin efficiency correction, obtained from the MC simulation.
Summing over all bins within the
interval $x(D_s) < 0.5$ and dividing by the $D_s^+$ and $\phi$ 
decay branching fractions
and by the number of $b\bar{b}$ events in the $\Upsilon$(5S) data sample,
the inclusive branching fraction
${\cal B}(\Upsilon{\rm (5S)}\,\to\,D_s X) / 2 = (23.6 \pm 1.2 \pm 3.6)\%$
is obtained.
In the calculations the PDG value 
${\cal B} (D_s^+\,\to\,\phi \pi^+) = (4.4 \pm 0.6)\%$ \cite{pdg}
is used.
The $\Upsilon$(5S) inclusive branching fraction is multiplied
by a factor of 1/2 to compare with $B_{(s)}$ branching fractions.
As explained above, continuum $b\bar{b}$ production
cannot be separated from $\Upsilon$(5S) events and therefore
continuum $b\bar{b}$ production is included in the
$\Upsilon$(5S)$\to D_s X$ branching fraction.
The latter is therefore defined as
the average number of $D_s$ mesons produced in $b\bar{b}$ 
events at the $\Upsilon$(5S) energy.

The dominant contributions to the
systematic uncertainty on the branching fraction measurement are 
the uncertainties due to the
${\cal B} (D_s^+\,\to\,\phi \pi^+)$ measurement of $14\%$, 
to the number of $b\bar{b}$ events of $5\%$,
to the track reconstruction efficiency and
particle identification of $4\%$, and to the fit procedure of $2\%$.

\begin{figure}[t!]
\epsfig{file=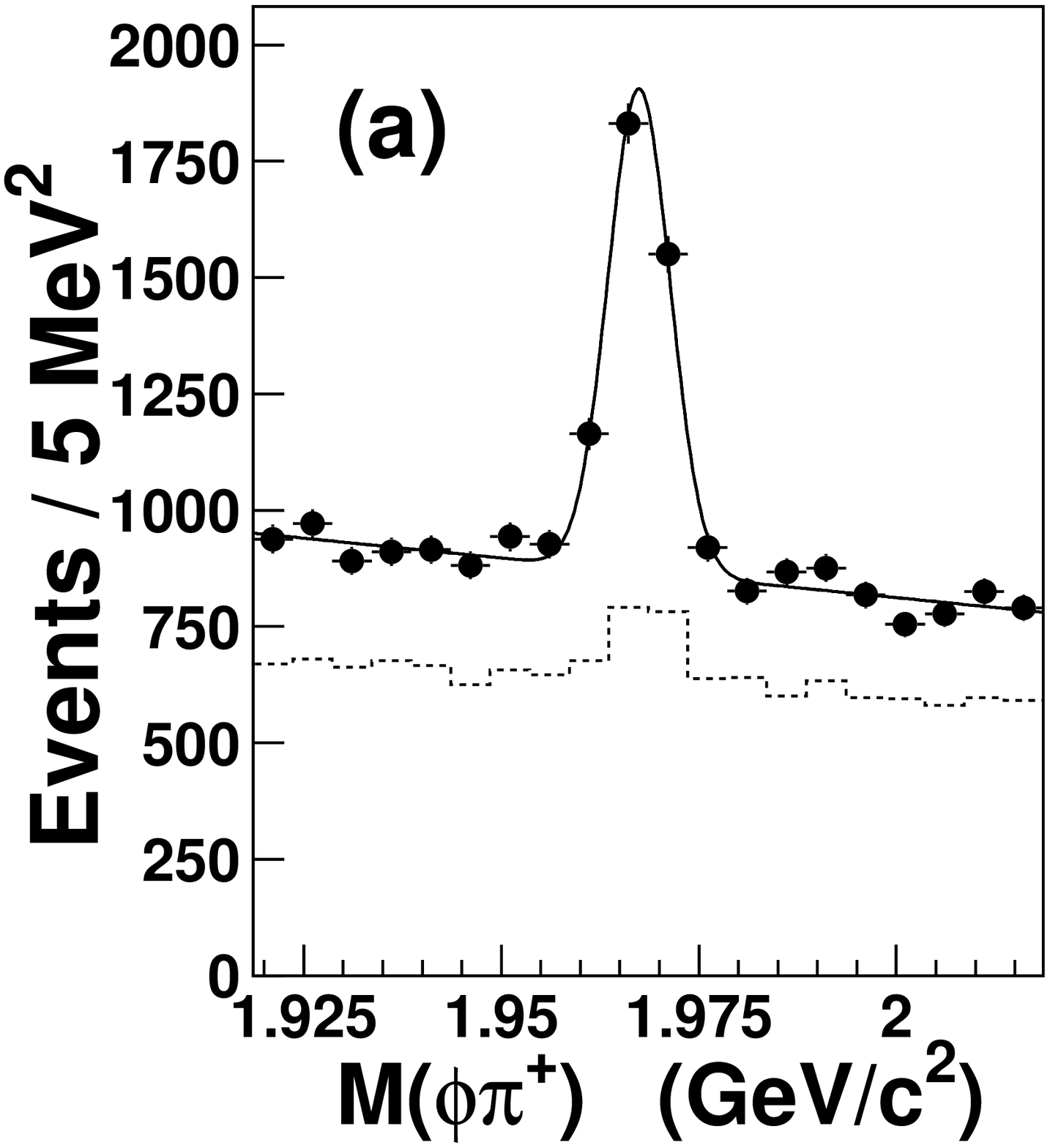,width=4.2cm,height=4.2cm}\epsfig{file=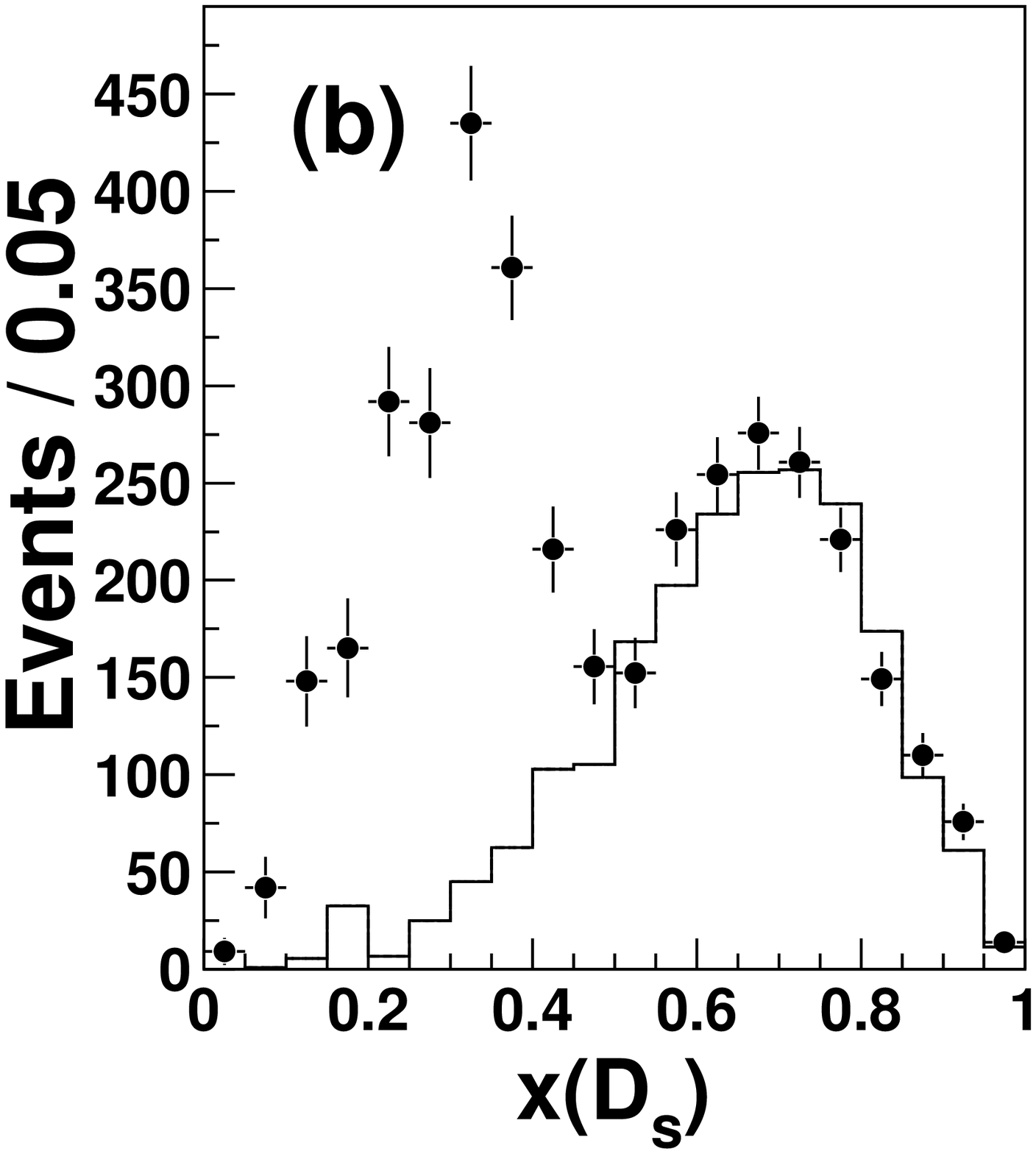,width=4.2cm,height=4.2cm}
\vspace{-0.2cm}
\caption{The $D_s$ signal in the region $x(D_s) < 0.5$ (a)
and the $D_s$ normalized momentum $x(D_s)$ (b).
The points with error bars are the
$\Upsilon$(5S) data, while the histograms show the 
normalized continuum (here and in Figs.~3 and 4).}
\vspace{-0.2cm}
\label{fig1}
\end{figure}

The obtained inclusive branching fraction
agrees well with the branching fraction
${\cal B}(\Upsilon{\rm (5S)}\,\to\,D_s X) / 2 = (22.4 \pm 2.1 \pm 5.0)\%$
obtained by CLEO \cite{cleoi}.
The value of ${\cal B}(\Upsilon{\rm (5S)}\,\to\,D_s X) / 2$
is significantly larger than the branching fraction 
${\cal B}(B\,\to\,D_s X) = (8.7 \pm 1.2)\%$,
which we calculate by combining the PDG average \cite{pdg} with the recent
CLEO result \cite{cleoi}
adjusted to the value ${\cal B} (D_s^+\,\to\,\phi \pi^+) = (4.4 \pm 0.6)\%$.
The significant increase of $D_s$ production at the
$\Upsilon$(5S) as compared to that at the $\Upsilon$(4S) indicates a sizable 
$B_s$ production rate.

The fraction $f_s$ of $B_s^{(\ast)} \bar{B}_s^{(\ast)}$ events among 
all $b\bar{b}$ events at the $\Upsilon$(5S) satisfies the 
following relation:
\begin{eqnarray}
\nonumber
& {\cal B}(\Upsilon{\rm (5S)}\,\to\,D_s X) / 2 = f_s \cdot {\cal B}(B_s\,\to\,D_s X)\ + \ \nonumber \\
& (1 - f_s) \cdot {\cal B}(B\,\to\,D_s X) \ \ ,
\end{eqnarray}
where ${\cal B}(B_s\to\,D_s X)$ and ${\cal B}(B\to\,D_s X)$ are 
the average number of $D_s$ mesons produced in $B_s$ and $B$ 
decays, respectively. Using our measurement of
${\cal B}(\Upsilon{\rm (5S)}\to\,D_s X)$, the measured value of
\mbox{${\cal B}(B\,\to\,D_s X) =$}
\mbox{$(8.7 \pm 1.2)\%$} \linebreak \mbox{\cite{pdg,cleoi}},
and the model-dependent estimate
${\cal B}(B_s\,\to\,D_s X) = (92 \pm 11)\%$ \cite{cleoi},
we determine $f_s = (17.9 \pm 1.4 \pm 4.1)\%$.
The systematic uncertainty on $f_s$ is obtained by propagating the
systematic uncertainties on the branching fractions included in Eq.\ (2),
taking into account the correlation induced by 
${\cal B} (D_s^+\,\to\,\phi \pi^+)$.

The inclusive production of $D^0$ mesons (including 
both $D^0$ and $\bar{D}^0$)
at the $\Upsilon$(5S) is
studied applying
a procedure similar to that used for the $D_s$ case.
As shown in \mbox{Fig.\ 3(a)}, large $D^0$ signals ($D^0\,\to\,K^- \pi^+$) 
are seen in the $\Upsilon$(5S) and continuum data samples.
The number of $D^0$ mesons as a function of $x(D^0)$
is shown in Fig.\ 3(b)
for both samples.
These two distributions agree well in the region $x(D^0) > 0.5$.

\begin{figure}[t!]
\epsfig{file=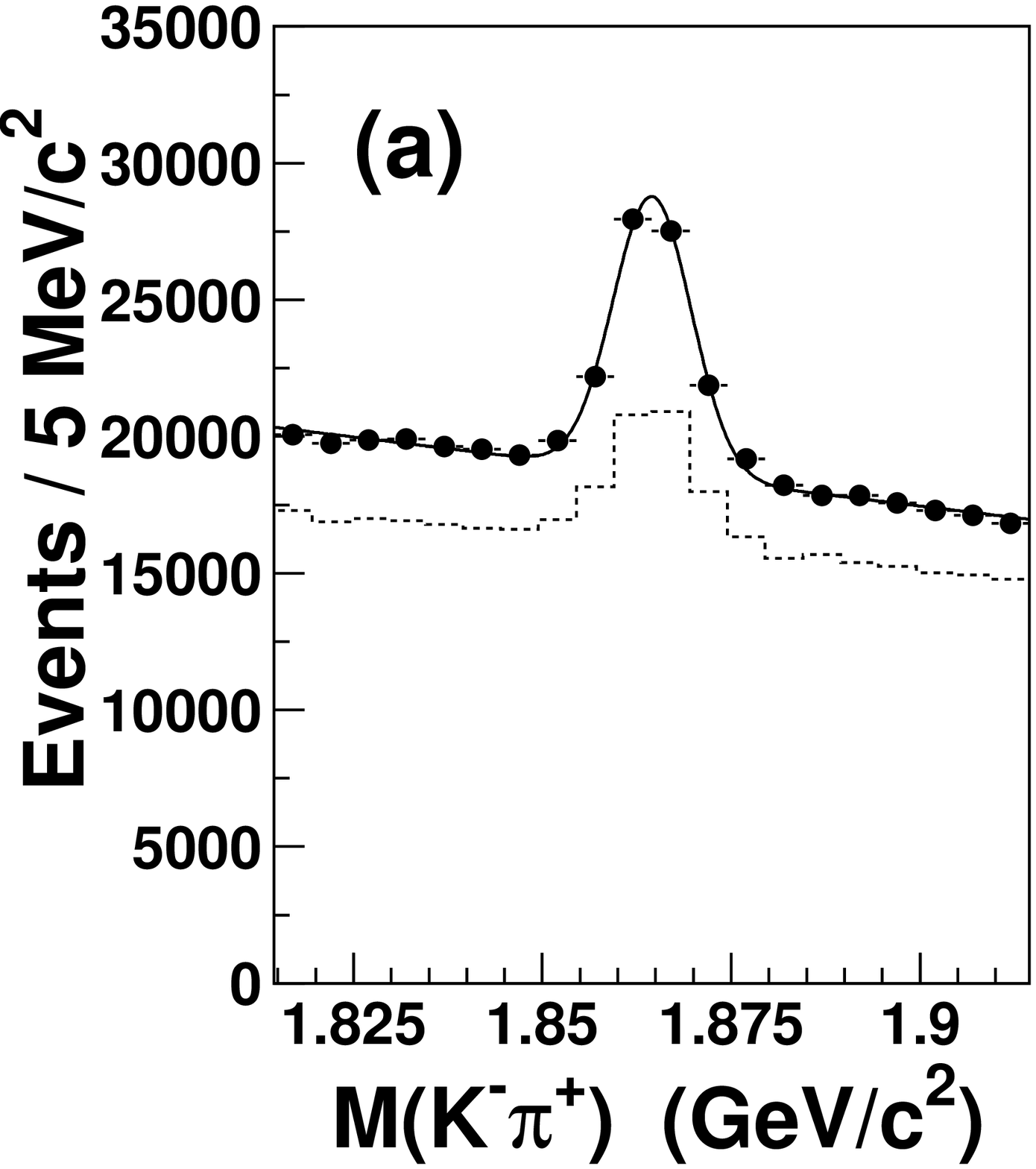,width=4.2cm,height=4.2cm}\epsfig{file=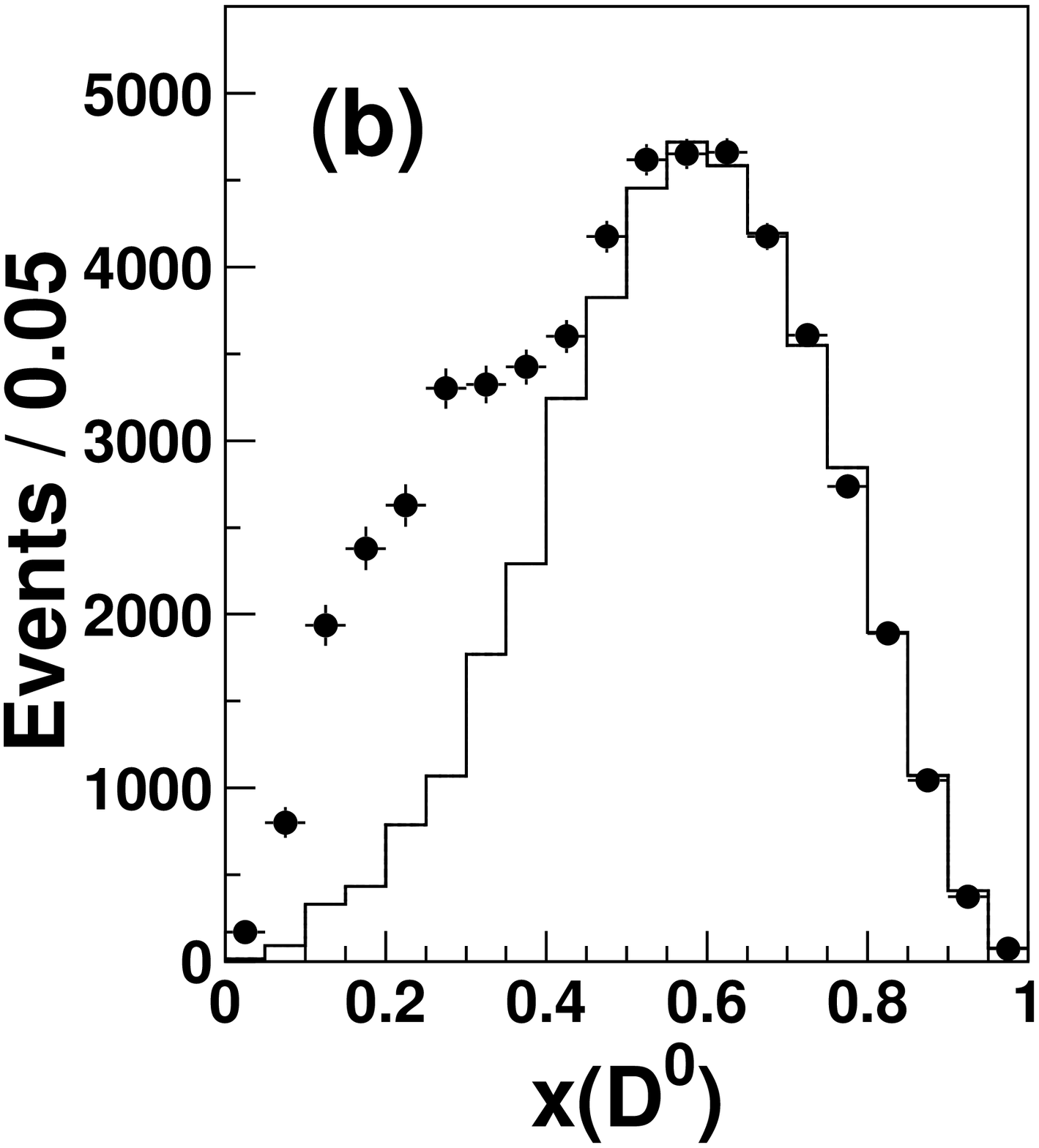,width=4.2cm,height=4.2cm}
\vspace{-0.2cm}
\caption{The $D^0$ signal in the region $x(D^0) < 0.5$ (a)
and the $D^0$ normalized momentum $x(D^0)$ (b).
}
\vspace{-0.2cm}
\label{fig2}
\end{figure}

After continuum subtraction and efficiency
correction, the inclusive branching fraction
${\cal B}(\Upsilon{\rm (5S)}\,\to\,D^0 X) / 2 = (53.8 \pm 2.0 \pm 3.4)\%$ is 
determined.
This branching fraction is defined as
the average number of $D^0$ and $\bar{D}^0$ mesons produced per $b\bar{b}$
event.
The dominant sources of systematic uncertainties are similar 
to the $D_s$ analysis, except that the PDG value
${\cal B} (D^0\,\to\,K^- \pi^+) = (3.80 \pm 0.07)\%$ has much
better accuracy than ${\cal B} (D_s^+\,\to\,\phi \pi^+)$.

The value obtained for 
${\cal B}(\Upsilon{\rm (5S)}\,\to\,D^0 X) / 2$
is lower than the PDG value
${\cal B}(B\,\to\,D^0 X) =$ \mbox{$(64.0 \pm 3.0)\%$} \cite{pdg},
as expected, if there is sizable $B_s$ production at the $\Upsilon$(5S).
The rate of $D^0$ production in $B_s$ decays can be estimated
in a way similar to that described in Ref.~\cite{cleoi}. Assuming  
that $D^0$ mesons are dominantly produced from conventional 
$b \to c$ processes through a fragmentation mechanism with 
additional $u\bar{u}$ quark pair creation, the branching fraction is expected 
to be ${\cal B}(B_s\,\to\,D^0 X) = (8 \pm 7)\%$.
Using the inclusive $D^0$ production branching fraction
of the $\Upsilon$(5S), $B$, and $B_s$ decays and replacing $D_s$ by
$D^0$ in Eq.\ (2),
the ratio $f_s = (18.1 \pm 3.6 \pm 7.5)\%$ of 
$B_s^{(\ast)} \bar{B}_s^{(\ast)}$ events to all $b\bar{b}$ events
at the $\Upsilon$(5S) is obtained.
The systematic error is dominated by
the systematic uncertainties from
${\cal B}(\Upsilon{\rm (5S)}\,\to\,D^0 X) / 2$ and ${\cal B}(B\,\to\,D^0 X)$
and is slightly affected by the uncertainty on the 
model-dependent assumption for the ${\cal B}(B_s\,\to\,D^0 X)$ value.
Although the $D_s$ inclusive analysis provides better accuracy on $f_s$, 
the $D^0$ inclusive analysis is an
independent method, where the uncertainty due to the $D^0$ decay 
branching fraction is small and the uncertainty on the number of
$b\bar{b}$ events dominates.
Moreover, the correlation between $f_s$ and the number of $b\bar{b}$ events
is positive in the $D^0$ analysis and is negative in the $D_s$ analysis.

Using the $f_s$ value obtained in the $D^0$ inclusive analysis,
the inclusive branching fraction ${\cal B}(B_s\,\to\,D_s X)$
can be extracted from Eq.\ (2).
Using the results of the present analysis, and taking into account 
the correlation between our measurements of 
${\cal B}(\Upsilon{\rm (5S)}\,\to\,D_s X)$ and ${\cal B}(\Upsilon{\rm (5S)}\,\to\,D^0 X)$,
we obtain ${\cal B}(B_s\,\to\,D_s X) = (91 \pm 18 \pm 41)\%$,
in agreement with expectations within large errors.

The inclusive production of $J/\psi$ mesons is
studied in the decay mode $J/\psi\,\to\,\mu^+ \mu^-$. 
As shown in Fig.\ 4(a),
the $\Upsilon$(5S) data sample contains a prominent $J/\psi$ signal, 
whereas the $J/\psi$ signal in continuum is small.
The $x(J/\psi)$ distributions are shown in Fig.\ 4(b)
for the $\Upsilon$(5S) and continuum data samples.
As expected, $J/\psi$ production in the continuum region
$x(J/\psi) > 0.5$ is smaller than in the low $x(J/\psi)$ region 
where $B$ and $B_s$ mesons can contribute.

\begin{figure}[t!]
\epsfig{file=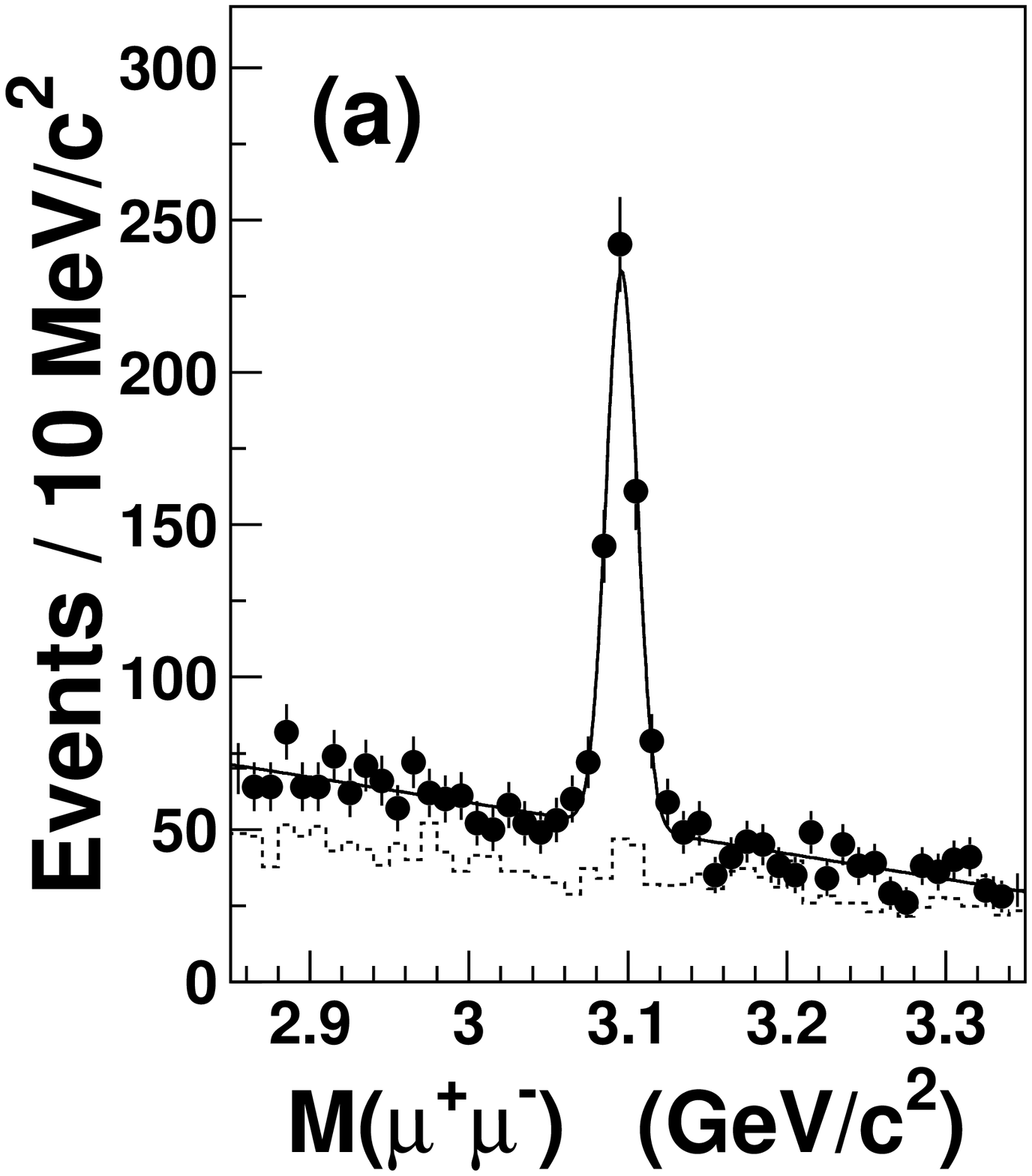,width=4.2cm,height=4.2cm}\epsfig{file=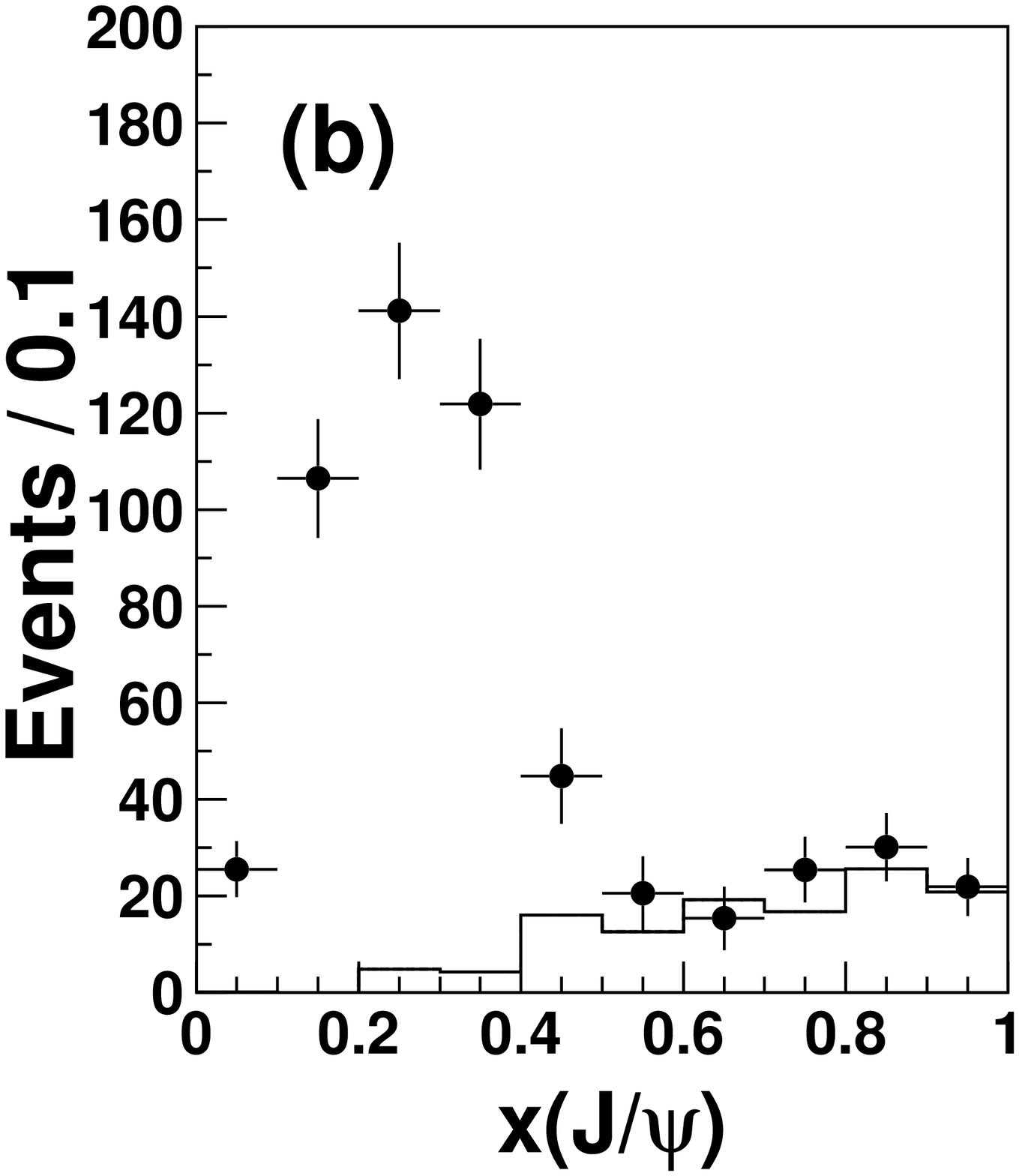,width=4.2cm,height=4.2cm}
\vspace{-0.2cm}
\caption{The $J/\psi$ signal in the region $x(J/\psi) < 0.5$ (a)
and the $J/\psi$ normalized momentum $x(J/\psi)$ (b).
}
\vspace{-0.2cm}
\label{fig3}
\end{figure}

The difference of these distributions in the region $x(J/\psi)\,<\,0.5$
is corrected for efficiency to
obtain the inclusive $J/\psi$ spectrum in $b\bar{b}$ events.
Using the PDG value
\mbox{${\cal B}(J/\psi\,\to\,\mu^+ \mu^-) =$}
$(5.88 \pm 0.10)\%$ \cite{pdg},
the inclusive branching fraction 
\mbox{${\cal B}(\Upsilon{\rm (5S)}\,\to\,J/\psi X) / 2 =$}
$(1.030 \pm 0.080 \pm 0.067)\%$
is obtained.
Systematic uncertainties are similar to those in the $D_s$ analysis
and are dominated by the uncertainty on the number of $b\bar{b}$ events.
The $\Upsilon$(5S) branching fraction can be compared with
${\cal B}(B\,\to\,J/\psi X) = (1.094 \pm 0.032)\%$ \cite{pdg},
because the inclusive $J/\psi$ production rates
in $B$ and $B_s$ decays are expected to be approximately equal.
Conversely, assuming the equality of inclusive $J/\psi$
branching fractions in $B$ and $B_s$ decays, we can obtain
the number of $b\bar{b}$ events. This cross-check can be important for 
future large statistics $\Upsilon$(5S) measurements.

In conlusion, the inclusive production of $D_s$, $D^0$, and $J/\psi$ mesons
has been studied in $e^+ e^-$ collisions at the $\Upsilon$(5S) energy.
The precise measurement of 
\mbox{${\cal B}(\Upsilon({\rm 5S})\rightarrow D_s X)$} and
the first measurement of 
\mbox{${\cal B}(\Upsilon({\rm 5S})\rightarrow D^0 X)$} and
\mbox{${\cal B}(\Upsilon({\rm 5S})\rightarrow J/\psi X)$}
branching fractions are performed. 
Clear evidence of sizable $B_s$ production is observed. 
The fraction of $B_s^{(\ast)} \bar{B}_s^{(\ast)}$ events among
all $b\bar{b}$ events produced at the $\Upsilon$(5S) energy is
determined from the inclusive $D_s$ and $D^0$ analyses
to be \mbox{$f_s = (17.9 \pm 1.4 \pm 4.1)\%$} and
\mbox{$f_s = (18.1 \pm 3.6 \pm 7.5)\%$}, respectively.
Combining these two $f_s$ measurements and taking into account the 
anticorrelated systematic uncertainty due to the number of $b\bar{b}$ 
events, the value $f_s =$ \mbox{$(18.0 \pm 1.3 \pm 3.2)\%$} is obtained.
This measurement agrees with the CLEO value of 
$f_s =$ \mbox{$(16.0 \pm 2.6 \pm 5.8)\%$} \cite{cleoi},
but has a total relative uncertainty that is a factor of two smaller.
The $B_s$ production rate over all $b\bar{b}$ events
at the $\Upsilon$(5S) is somewhat larger than the fraction of $B_s$
mesons produced from the $\bar{b}$ quark in 
$Z \rightarrow b\bar{b}$ decays at LEP,
${\cal B}(\bar{b} \rightarrow B_s) = (10.2 \pm 0.9) \%$ \cite{pdg}.
The large $f_s$ value measured here indicates very good potential
for future $B_s$ studies at high luminosity asymmetric-energy $B$ factories
running at the $\Upsilon$(5S) resonance.

We thank the KEKB group for excellent operation of the
accelerator, the KEK cryogenics group for efficient solenoid
operations, and the KEK computer group and
the NII for valuable computing and Super-SINET network
support.  We acknowledge support from MEXT and JSPS (Japan);
ARC and DEST (Australia); NSFC and KIP of CAS (China); 
DST (India); MOEHRD, KOSEF and KRF (Korea); 
KBN (Poland); MIST (Russia); ARRS (Slovenia); SNSF (Switzerland); 
NSC and MOE (Taiwan); and DOE (USA).

\end{document}